\documentclass[twocolumn,nofootinbib,amsmath,amssymb,a4paper]{revtex4}
\usepackage{graphicx}
\usepackage{dcolumn}
\usepackage{bm}

\newcommand{\beq}{\begin{equation}}
\newcommand{\eeq}{\end{equation}}

\newcommand{\nn}{\nonumber}
\newcommand{\vtstd}{V^*_{ts}V^{\phantom{*}}_{td}}
\newcommand{\vckm}{{V_{\text{CKM}}}}
\newcommand{\texthalf}{{\textstyle{\frac12}}} 
\newcommand{\LLN}{{\Lambda_{LN}}}
\newcommand{\hc}{\text{h.c.}}

\newcommand{\yten}{\lambda_{10} }

\newcommand{\ynu}{\lambda_{1}}
\newcommand{\yfiveph}{\lambda_{5}^{\phantom{\dagger}} }
\newcommand{\ytenph}{\lambda_{10}^{\phantom{\dagger}} }
\newcommand{\ysigph}{\lambda_{5}^{\prime\phantom{\dagger}} }
\newcommand{\ysfph}{\lambda_{5}^{(\prime)\phantom{\dagger}}}
\newcommand{\ynuph}{\lambda_{1}^{\phantom{\dagger}}}

\newcommand{\yuu}{\lambda_u}
\newcommand{\ydd}{\lambda_d}
\newcommand{\yee}{\lambda_e}
\newcommand{\ynn}{\ynu}
\newcommand{\yuuph}{\lambda_u^{\phantom{\dagger}}}
\newcommand{\yddph}{\lambda_d^{\phantom{\dagger}}}
\newcommand{\yeeph}{\lambda_e^{\phantom{\dagger}}}
\newcommand{\ynnph}{\ynuph}

\begin{document}

\title{Minimal Flavor Violation}

\author{Benjam\'\i n Grinstein}
 \email{bgrinstein@ucsd.edu}
\affiliation{%
Physics Department, University of California, San Diego; La Jolla, CA
 92093-0319, USA
}%

\begin{abstract}
If new physics is called upon to explain away fine tunings, like the
hierarchy problem, then, we argue, the principle of Minimal Flavor
Violation is inescapable. We review the principle and recent extensions
to the lepton sector and to Grand-Unified theories.
\end{abstract}

\maketitle

\section{Introduction}
In the absence of new dynamics radiative corrections would render the
mass scale of the electroweak theory comparable to the Planck
scale. New physics at the TeV scale is generally invoked to explain
this {\it hierarchy problem.}  But quark mass terms break the
electroweak symmetry, so the quark mass matrices are necessarily
connected to this new physics. New higgs dynamics at the TeV scale
leads inescapably to new flavor physics. This statement is
straightforwardly verified in all available examples. Perhaps the most
popular, the MSSM, is the most obvious: the solution to the hierarchy
problem comes with a plethora of new fields carrying flavor, namely,
the squarks. But other, less well known examples, like the Lee-Wick
Standard Model\cite{Grinstein:2007mp}, similarly contain an intricate
flavor structure (in this case, an unstable resonance is associated
with each known particle).

To describe the effects of new TeV dynamics at below TeV energies in a
model independent approach one simply extends the Lagrangian of the
standard model (SM) by operators of dimension higher than four,
suppressed by powers of the new physics scale, $\Lambda$. Buchmuller and
Wyler\cite{Buchmuller:1985jz} and Leung, Love and
Rao\cite{Leung:1984ni},  listed all operators of dimension five and six
and analyzed their effects. Ignoring operators mediating flavor
changing neutral currents (FCNC), $\Lambda\sim$ a few TeV is consistent with
experiment. But if the coefficient of FCNC operators is given by
dimensional analysis, then $\Lambda\sim$ a few TeV is strongly excluded. A much
larger scale, $\Lambda~10^4$~TeV, is still consistent with experiment, but
then a hierarchy problem reappears.

So if we want to insist that the scale of new physics be a few TeV we
need some principle that will make the coefficient of the dangerous
FCNC operators automatically (naturally) small. The principle of
Minimal Flavor Violation (MFV) does just that. We will describe MFV
below, but we pause here to note the generality of these observations:
we have assumed that below a scale of about a TeV our model has the
field content of the standard model, and have insisted in the absence
of fine tunings (the very reason we need new physics at the TeV
scale). Moreover, the principle can be readily generalized to the
cases where below the scale of new physics the model
has two higgs doublets, or no higgs at all (a strongly coupled higgs
sector). It is this generality that I find so compelling. Not only
seems MFV inescapable, it seems we have learned something deep and
fundamental, namely, that the origin of flavor is to be found in some
secluded sector that expresses itself only through a single channel.

\section{MFV}
In the absence of quark masses the SM lagrangian has a
large exact flavor symmetry group, $G_F=SU(3)^3\times U(1)^2$, arising from
independent unitary rotations of the three flavors of quark doublets
$q_L$, and singlets $u_R$ and $d_R$. In the SM this  symmetry  is
broken only by the Yukawa terms in the lagrangian that result in quark
masses once the higgs gets a vev,
\begin{equation}
\label{yuks}
\mathcal{L}_Y=\lambda_U^{ij}H\bar q_L^iu_R^j+\lambda_D^{ij}\bar
q_L^id_R^j+\text{h.c.}
\end{equation}

The basic premise of the MFV hypothesis is that {\it there is a unique
source of breaking of the $G_F$ symmetry.}  We already have $G_F$
breaking in \eqref{yuks}, so any additional terms that break $G_F$
must transform under $G_F$ in exactly the same way as 
\eqref{yuks}. This principle can be implemented in extensions of the
standard model that incorporate new (yet undiscovered) fields that
carry flavor quantum numbers. But we will be interested in a model
independent analysis and this is accomplished by adjoining to the
SM all operators of dimension higher than four constructed of SM fields.
Those operators that break
$G_F$ must transform just as the Yukawa terms.

This all sounds very general and abstract. It is perhaps easier to
understand this in a particular context, so let's consider an
example. In the SM the flavor changing neutral current first appears
at one loop and is dominated by the graph with a top quark. The low
energy effective interaction hamiltonian for $K_L\to\pi\nu\bar\nu$ is
\begin{equation}
\mathcal{H}_{\rm eff,SM}=\frac{4G_F}{\sqrt2}~\mathcal{C}\! \sum_{\ell=e,\mu\tau}\!\! 
\bar s_L\gamma_\mu d_L\,\bar \nu_L^\ell\gamma^\mu \nu_L^\ell+\text{h.c.}
\end{equation}
where
\begin{equation}
\mathcal{C}=\left[\frac{\alpha }{2\pi\sin^2\theta_W}\,X(m_t/m_W)\right]\vtstd
\end{equation}
The factor in the square bracket includes the obvious electroweak
coupling constants and a function $X(x)$, with $X\sim1$ for $x\gtrsim 1$, that
results from performing the 1-loop integral. The second factor,
involving the product of CKM elements, 
contains the flavor information and makes the coefficient
$\mathcal{C}$ small. Recall,  the Wolfenstein parametrization
\beq
\vckm \approx \begin{pmatrix}
1-\texthalf\lambda^2&\lambda&A\lambda^3(\bar\rho-i\bar\eta)\\
-\lambda(1+iA^2\lambda^4\bar\eta) & 1-\texthalf\lambda^2& A\lambda^2\\
 A\lambda^3(1-\bar\rho-i\bar\eta)& -A\lambda^2(1+i\lambda^2\bar\eta)& 1
\end{pmatrix}\nn
\eeq
which is an expansion in the small parameter $\lambda\approx0.22$, the sine of the
Cabibbo angle. Note that $\vtstd\sim A^2\lambda^5$, that is, of fifth order in
the small parameter. Hence the Branching fraction for this process is
suppressed by $\sim A^4\lambda^8$.

Now consider the effects of new physics parametrized by dimension six
operators suppressed by  the new physics scale $\Lambda$,
\begin{equation}
\label{exampleNP}
\mathcal{H}_{\rm eff,NP}=\frac{1}{\Lambda^2}
\sum_{\ell=e,\mu\tau}\!\!\mathcal{C}^\ell_{\rm NP} \,
\bar s_L\gamma_\mu d_L\,\bar \nu_L^\ell\gamma^\mu \nu_L^\ell+\text{h.c.}
\end{equation}
Other dimension six operators can be added, but we consider one that
is identical to the operator that results from integrating out the
top-quark and $W$-boson in the SM so we may compare coefficients
directly. In the absence of fine tuning the the coefficients are
expected to be  order unity,   $\mathcal{C}^\ell\sim1 $ .

Now imagine an experiment is performed that has sensitivity to
a fractional deviation $r$ form the SM expectation. Then 
\beq
\label{KOPIOnoMFV}
1+r\sim\left|1+\frac{1/\Lambda^2}{A^2\lambda^5/(16\pi^2M_W^2)}\right|^2
\eeq
tells us the scale $\Lambda$ to which this experiment is
sensitive. For example, $r=4\%$, roughly the would be sensitivity of
the now canceled KOPIO experiment, translates into a reach of
$\Lambda\sim10^6$~GeV. Similarly, using measured FCNC processes, such as
 $K^0$-$\bar K^0$,   $B^0$-$\bar B^0$ or $B_s^0$-$\bar B_s^0$ mixing
or $B^0\to K^{0*}\gamma $ gives bounds on $\Lambda$ or order $10^6$~GeV.

The large suppression of the SM rate, of order $A^2\lambda^5$, arises from a
generalized GIM mechanism (suppressions from either small masses or
small mixing angles). As we will see, MFV guarantees that the same CKM
factor appears also in the new physics operator. The $A^2\lambda^5$ cancels
out in the ratio, so the estimate in \eqref{KOPIOnoMFV} is modified
under the MFV hypothesis to 
\beq
1+r\sim\left|1+\frac{1/\Lambda^2}{1/(16\pi^2M_W^2)}\right|^2.
\eeq
Now $r=4\%$ gives $\Lambda~10^{3-4}$~GeV. A comprehensive, detailed analysis
of bounds on $\Lambda$ an be found in \cite{D'Ambrosio:2002ex}.

So the only thing left to understand in our example is how the MFV
principle inserts automatically a factor of $A^2\lambda^5$ into the coefficient
of the new physics operator in Eq.~\eqref{exampleNP}. It is
straightforward to implement the MFV principle using the spurion
method. The SM lagrangian is invariant under the following combined
transformation of fields an couplings:
\begin{equation}
\label{spurionMFV}
\begin{aligned}
q_L&\to V_L q_L \\
u_R&\to V_u u_R \\
d_R&\to V_d d_R \\
\end{aligned}
\qquad\qquad
\begin{aligned}
\lambda_U&\to V_L \lambda_U V_u^\dagger\\
\lambda_D&\to V_L \lambda_D V_d^\dagger
\end{aligned}
\end{equation}
This is a  $G_F$ transformation if the fields and  it would be a
symmetry of the theory if $\lambda_U=\lambda_D=0$. So the transformation of the
matrices $\lambda_U$ and $\lambda_D$ characterize the breaking of $G_F$. To
implement the MFV principle we simply need to insist that our
modifications to the theory preserve the invariance under
\eqref{spurionMFV}. 

Consider the operator in the effective hamiltonian  of our example,
Eq.~\eqref{exampleNP}. The quark fields are components of the $q_L$
flavor triplet and as written the operator is not invariant
under~\eqref{spurionMFV}. To fix this replace the quark bilinear $\bar
s_L\gamma_\mu d_L $ by 
\beq
 \bar q_L \lambda_U^{\phantom{\dagger}} \lambda_U^\dagger \gamma_\mu q_L 
\to \left(\sum_{x=u,c,t}V_{xs}^*V_{xd}^{\phantom{*}}\frac{m_q^2}{v^2}\right)\bar s_L\gamma_\mu d_L 
\eeq 
where in the last step  we have indicated the $\Delta S=1$ piece in the
mass eigenstate basis. The dominant term in the sum is from $x=t$ and
gives $ \vtstd m_t^2/v^2\approx A^2\lambda^5$. 

\subsection{Simple extensions}
The analysis presented above is model independent only to a point: we
assumed that below the scale of new physics, $\Lambda$, the spectrum
is that of the SM with a single higgs doublet. The analysis has to be
modified if this is not the case. An interesting example is that of
the SM with two higgs doublets. MFV requires that the
Yukawa couplings of the two higgs doublets to quarks be restricted
since there can only be two truly independent, fundamental matrices
that break $G_F$. In the generic case, FCNC  appear from tree level exchange of
neutral higgs particles. There are also new radiative contributions to
FCNC from charge higgs exchange. Hence some couplings have to be
 restricted further. This is accomplished naturally by assuming approximate
Peccei-Quinn (PQ) symmetry, and that the PQ symmetry violating terms are
controlled by a new small parameter.

Even then, the two higgs doublet model has more parameters than the
one higgs SM: the ratio of expectation values of the two higss
doublets $v_2/v_1\equiv\tan\beta$ and the masses of three additional scalar
particles (one charged and two neutral).  Some of the interest in
these models is from possibly describing the hierarchy of the top and
bottom masses by a hierarchy in expectations values, $\tan\beta \gg1$. Since
this requires larger $\lambda_D$ than in the single higgs SM, FCNC are
enhanced. 

The analysis of the effects of higher dimension operators of the two
higgs model is then similar to that of the one higgs case, with two
important distinctions. (i) Coefficients of operators involving down
type masses are enhanced by corresponding powers of $\tan\beta$, and
(ii) There are additional contributions to FCNC mediated by the
additional fields, e.g, charged higgs exchange.

\section{Leptons: MLFV}
We do not know why but MFV seems to be operative in the quark
sector. Surely we need more work to establish that this is accurately
true. But in the mean time it is clearly interesting to ask if MFV is
a more general principle. If so, we wonder, shouldn't it also apply
to the lepton sector of the SM? Lepton flavor is violated, as
evidenced by  neutrino oscillations. It is interesting to investigate
if Minimal Lepton Flavor Violation (MLFV) makes interesting
predictions of flavor lepton changing neutral currents of charged
lepton, e.g., $\mu\to e\gamma$ and $\mu\to ee\bar e$. In particular, we may ask not
just about the magnitude of these effects but more particularly
whether there are particular patterns of flavor violation that may
help us decide if indeed MLFV is the underlying structure. If this
were the case it would strengthen the notion that MFV operates at a
very basic level

There are two cases to consider\cite{Cirigliano:2005ck}. If the
neutrinos acquire dirac masses the analysis of flavor changing neutral
currents proceeds in exactly the same way as for the quark sector. In
this case the tiny neutrino mass makes all charged lepton FCNC
impossibly small to observe, and for that reason we do not pursue this
further. The situation is very different if the neutrinos acquire a
majorana mass. A majorana mass is attractive in any case because it
can explain the smallness of the neutrino masses through the
``see-saw'' mechanism.

The analysis of MLFV for neutrinos with majorana mass does not require
that we add right handed neutrinos to generate masses: an operator of
dimension 5 (see below, Eq.~\eqref{minFC}) can produce the desired
see-saw mass. Therefore we examine two cases:
\begin{enumerate}
\item {\it Minimal field content (MFC)}: the same leptonic field content as
the SM: three left-handed and lepton doublets $L_{L}^i$ and three
right-handed charged lepton singlets $e_{R}^i$ In this case the lepton
flavor symmetry group is $ G_{\rm LF} = SU(3)_L\times SU(3)_E~.  $ The
lepton sector is also invariant under two $U(1)$ symmetries, which can
be identified with the total lepton number, $U(1)_{\rm LN}$, and the
weak hypercharge.
\item {\it Extended field content (EFC)}: three 
right-handed neutrinos, $\nu^i_R$, in addition to the SM fields.
In this case the field content of the 
lepton sector is very similar to that of the quark sector, 
with a maximal flavor group $G_{\rm LF} \times SU(3)_{\nu_R}$.
\end{enumerate}
This classification is very reasonable if the masses of right handed
neutrinos, $M_R$, are smaller than the scale of new EW-physics, $\Lambda$. But
really what we have in mind is the extreme opposite case, since we
want roughly $m_\nu\sim v^2/M_R\sim (\Lambda/4\pi)^2/M_R$. By
considering the two cases we can examine the difference that arise by
assuming that the parameter that controls MLFV is the coefficient of a
dimension 5 operator (MFC) or a Yukawa interaction (EFC). However, due
to space constraints I will describe here only the MFC and refer the
reader to Ref.~\cite{Cirigliano:2005ck} for a more detailed description of the EFC.

\subsection{Minimal Field Content} We  make three assumptions: (i)The breaking of
the $U(1)_{\rm LN}$ is independent from the breaking of the lepton
flavor symmetry ($G_{\rm LF}$), (ii) The breaking of $U(1)_{\rm LN}$
is associated with a very high scale $\LLN$, much greater than the
scale of EW physics, $\LLN\gg\Lambda$, and (iii)There are only two
irreducible sources of lepton-flavor symmetry breaking,
$\lambda_e^{ij}$ and $g_\nu^{ij}$, defined by
\begin{align}
\label{minFC}
\mathcal{H}_{\rm eff}&= \lambda_e^{ij} \,\bar e^i_R(H^\dagger L^j_L) +\frac1{2\LLN}\,g_\nu^{ij}(\bar
L^{ci}_L\tau_2 H)(H^T\tau_2L^j_L)+\hc\nn\\
&\to v \lambda_e^{ij} \,\bar e^i_Re^j_L
+\frac{v^2}{\LLN}\,g_\nu^{ij}\,\bar\nu^{ci}_L\nu^j_L+\hc
\end{align}
In the second line we have indicated the mass terms after shifting the
higgs field by its expectation value. It displays explicitly the
see-saw mechanism. The scale $\LLN$ rather than $\Lambda$ appears in
the second terms because the operator breaks $U(1)_{\rm LN}$. 

The principle of MLFV with MFC can be implemented much like MFV, using
the spurion method. The hamiltonian~\eqref{minFC} is formally
invariant under the combined transformation lepton fields and the
matrices $\lambda_e^{ij}$ and $g_\nu^{ij}$ under $G_{\rm LF}$ are
\begin{equation}
\begin{aligned}
L_L &\to  V_L \,L_L\\
e_R &\to  V_R \,e_R
\end{aligned}
\qquad\qquad
\begin{aligned}
\lambda_e &\to V_R^{\phantom{\dagger}} \,\lambda_e V_L^\dagger~, \\
g_\nu &\to V_L^{*\phantom{\dagger}} g_\nu V_L^\dagger~.
\end{aligned}
\end{equation}
 It is readily seen that the quantity $\Delta\equiv g^\dagger_\nu g^{\phantom{\dagger}}_\nu$
has a simple transformation law and largely controls all FCNC of
charged leptons (a few four lepton operators involve also the
parameter $\delta\equiv g_\nu$\cite{Cirigliano:2006su}). Moreover, up to a constant it is determined by
quantities that are measurable at low energies; in the mass eigenstate
basis, $\Delta=(\LLN/v^2)^2Um_\nu^2U^\dagger$ (and $\delta =
(\LLN/v^2)U^*m_\nu U^\dagger$), where $U$ is the PMNS matrix and $m_\nu$ is the
diagonal neutrino mass matrix. Hence, all FCNC amplitudes are given in
terms of\cite{Cirigliano:2005ck,Cirigliano:2006su} (i)the ratio $\LLN/ \Lambda$ (but not both scales
independently), (ii)a few operator coefficients of order 1, and
(iii)low energy measurable (or measured) neutrino masses and mixing
angles. As a result this setup is very predictive. In some cases, like
the three radiative decays, $\mu\to e\gamma$, $\tau\to e\gamma$ and $\tau\to \mu
\gamma$, the
unknown parameters completely drop out of ratios of Branching
fractions. So this scheme is falsifiable. 

\section{MFV and GUT}
Applying the principle of MFV to Grand Unified Theories (GUTs)
produces interesting predictions. This has been noted in the
particular case of supersymmetric GUTs\cite{Barbieri:1994pv} but
similar
predictions occur more generally\cite{Grinstein:2006cg}, as described bellow.

For definiteness consider GUTs with $SU(5)$ as gauge group. The 15
fields of one family of quarks and leptons fall into a $\psi\sim\bar5$ and
a $\chi\sim10$ representations. Since $\psi$ and $\chi$ contain both
leptons and quarks, the flavor symmetry group of the GUT, $SU(3)^2$, is smaller
than in the SM for three generations.  As in the case of
MFV, we assume that the flavor symmetry is broken only by three
Yukawa-like couplings, the ones responsible for quark and lepton
masses:
\begin{equation}
{\cal L}_{\rm sym.br.}=\psi^T\lambda_5\chi H^*+\chi^T\lambda_{10}\chi H+\frac1M\psi^T\lambda'_{5}\Sigma\chi H^*
\end{equation}
The first trans-Planckian correction has been included. This is
necessary to accommodate the masses of all quarks and charged
leptons. The effects of this term are small, very naturally
accommodating the observed spectrum. Neutrinos masses can also be
included by adding three right handed neutral fields, $N$. A large Majorana
mass $M_R$ for these, 
\begin{equation}
\Delta{\cal L}_{\rm sym.br.}=N^T\lambda_1\psi H+N^TM_RN
\end{equation}
produces small, see-saw Majorana masses for the left handed neutrinos:
With these additional fields the flavor group, $SU(3)^3$, is larger.
So the MFV hypothesis applied to GUTs is the statement that the
$SU(3)^3$ flavor symmetry is broken only by the couplings
$\lambda_5,\lambda_5',\lambda_{10},\lambda_1,M_R$. It is convenient to
trade the parameters $\lambda_5,\lambda_5',\lambda_{10}$ for the low
energy combinations that give masses to quarks and leptons, since
those are of direct phenomenological relevance,
$  \yuu \sim  \ytenph$, $  \ydd\sim  (\yfiveph +  \ysigph)$, $\yee^T \sim (\yfiveph - \frac{3}{2}  \ysigph)$.

As before, it is
simplest to implement MFV in GUTs by the spurion method. the
transformation rules are:
\begin{equation}
\begin{aligned}
Q_L &\to V_{10}^{\phantom{*}}   ~Q_L \\
u_R &\to V_{10}^* ~u_R \\
d_R &\to V_{\bar 5}^* ~d_R
\end{aligned}
\qquad\quad
\begin{aligned}
\yuu  &\to  V_{10}^{*\phantom{\dagger}}     ~\yuuph~ V_{10}^\dagger  \\
\ydd &\to  V_{\bar 5}^{*\phantom{\dagger}} ~\yddph~ V_{10}^\dagger \\ 
\yee  &\to  V_{\bar 5}^{*\phantom{\dagger}} ~\yeeph~ V_{10}^\dagger 
\end{aligned}
\nonumber
\end{equation}
\begin{equation}
\begin{aligned}
L_L &\to V_{\bar 5}^{\phantom{*}}    ~L_L \\
e_R &\to V_{10}^* ~e_R 
\end{aligned}
\qquad\quad
\begin{aligned}
\ynu   &\to  V_{1}^{*\phantom{\dagger}}   ~\ynuph~  V_{\bar 5}^\dagger  \\
 M_R   &\to   ~V_{1}^{*\phantom{\dagger}}  ~M_R^{\phantom{\dagger}}~ V_{1}^\dagger  
\end{aligned}
\end{equation}
As a result of the reduced flavor symmetry, quarks and leptons
transform together. So in addition to some of the older bilinear
building blocks we encountered before, like
$ {\bar Q}_L \yuu^\dagger \yuuph Q_L,$
one encounters bilinear invariants that mix quark and lepton
 parameters, like 
${\bar Q}_L (\yeeph \yee^\dagger)^T Q_L $ and 
$\bar L_L(\yddph \ydd^\dagger )^T L_L$,
where it is understood that the substitution $
\lambda_e\leftrightarrow\lambda_d^T $ can be made throughout. There
are also new interesting leptoquark bilinears that are allowed. 

The phenomenology of these models is quite reach. The bottom line is
the inescapable appearance of lepton flavor changing interaction of
charged leptons, much as in the case of MLFV  above, but now with a
richer source of flavor violation.  For example the radiative decays  $\mu\to e\gamma$, $\tau\to e\gamma$ and $\tau\to \mu
\gamma$, are mediated by the low energy effective lagrangian
\begin{align*}
 {\cal L}_{\rm eff}
=\frac{v}{\Lambda^2} \bar e_R \left[c_{1}\yeeph \ynn^\dagger\ynnph 
+c_{2}\yuuph\yuu^\dagger \lambda_e
+c_{3}\yuuph\yuu^\dagger\lambda_d^T\right] \sigma^{\mu \nu} e_L F_{\mu \nu }
\end{align*}
This is more general than the SUSY-GUT result, so it is less specific
in its predictions. However there are several interesting aspects to
this results. First, depending on the parameters, this could be
dominated by the first MLFV-like terms, or by the the SUSY-GUT like
term, or neither. Secondly, the result is still hierarchical, as it
was in the MLFV case. If the second and third terms dominate, then the
branching amplitudes for  $\mu\to e\gamma$, $\tau\to e\gamma$ and $\tau\to \mu\gamma$ scale as
$\lambda^5m_\mu: \lambda^3m_\tau : \lambda^2 m_\tau $, where the mixing parameter is
$\lambda\approx0.22$. And thirdly, the rate is typically large, with the branching
fraction for $\mu\to e\gamma$ of order $10^{-12}$ for $\Lambda=10$~TeV.

\begin{acknowledgments}
Work supported in part by the Department of Energy under contract DE-FG03-97ER40546.
\end{acknowledgments}

\end{document}